\documentclass[noshowpacs,amsmath,
twocolumn,
superscriptaddress,
8pt
]{revtex4-1}
\usepackage[utf8]{inputenc}
\usepackage{setspace}
\usepackage{amsmath}
\usepackage{float}
\usepackage{bm}
\usepackage{ulem}
\usepackage{graphicx}
\usepackage[nearskip,margin = 0pt]{subfig}

\usepackage{verbatim}
\usepackage{amsfonts}
\usepackage{braket}
\usepackage{siunitx}
\usepackage{amssymb}
\usepackage{upgreek}
\usepackage[colorlinks,linkcolor=blue,anchorcolor=blue,citecolor=blue,urlcolor=black]{hyperref}
\usepackage{epstopdf}
\usepackage{xcolor}
\usepackage{booktabs}
\usepackage{tabularx}
\usepackage{xtab}
\usepackage{changepage}
\usepackage{ragged2e}

\DeclareGraphicsExtensions{.pdf,.eps,.png,.jpg,.mps} 

\begin{document}

\title{Layer-parity-defined surface polarization in Nb$_3$Cl$_8$ for excitonic modulation at van der Waals interfaces}
\author{Xinyue Huang$^{1,2,\dagger}$, Hansheng Xu$^{1,\dagger}$, Yuchen Gao$^{1,\star}$, Yushen Zhou$^{1}$, Zhijie Ma$^{3,4}$, Kenji Watanabe$^{5}$, Takashi Taniguchi$^{6}$, Zuxin Chen$^{7}$, Jianqi Huang$^{8,\star}$, Jianpeng Liu$^{8,9}$, Teng Yang$^{8,10,11}$, Youguo Shi$^{3,4,12}$, and Yu Ye$^{1,8,\star}$\\
\vspace{6pt}
$^1$State Key Laboratory for Mesoscopic Physics and Frontiers Science Center for Nano-optoelectronics, School of Physics, Peking University, Beijing 100871, China\\
$^2$Academy for Advanced Interdisciplinary Studies, Peking University, Beijing 100871, China\\
$^3$Beijing National Laboratory for Condensed Matter Physics and Institute of Physics, Chinese Academy of Sciences, Beijing 100190, China\\
$^4$University of Chinese Academy of Sciences, Beijing 100049, China\\
$^5$Research Center for Electronic and Optical Materials, National Institute for Materials Science, 1-1 Namiki, Tsukuba 305-0044, Japan\\
$^6$Research Center for Materials Nanoarchitectonics, National Institute for Materials Science, 1-1 Namiki, Tsukuba 305-0044, Japan\\
$^7$School of Semiconductor Science and Technology, South China Normal University, Foshan 528225, China\\
$^{8}$Liaoning Academy of Materials, Shenyang 110167, China\\
$^{9}$School of Physical Science and Technology, ShanghaiTech University, Shanghai 201210, China\\
$^{10}$Shenyang National Laboratory for Materials Science, Institute of Metal Research, Chinese Academy of Sciences, Shenyang 110016, China\\
$^{11}$School of Material Science and Engineering, University of Science and Technology of China, Shenyang 110016, China\\
$^{12}$Songshan Lake Materials Laboratory, Dongguan, Guangdong 523808, China\\
\vspace{3pt}
$^{\dagger}$These authors contributed equally to this work.\\
$^{\star}$Corresponding to: gaoyuchen@pku.edu.cn; jqhuang@lam.ln.cn; ye\_yu@pku.edu.cn}

\begin{abstract}
\begin{adjustwidth}{-2cm}{0cm}
\textbf{ABSTRACT:}
The intrinsic symmetry breaking in the breathing kagome lattice of layered Nb$_3$Cl$_8$ provides a unique mechanism for realizing electrically polar surfaces. In each monolayer, the trimerization of Nb atoms breaks inversion and mirror symmetries, generating an out-of-plane electric dipole. The AB-stacked $\alpha$ phase arranges adjacent layer dipoles antiferroelectrically, leaving the uncompensated surface polarization strictly governed by layer parity. Here, using atomic force microscopy operated in Kelvin probe force microscopy mode, we directly visualize layer-dependent polarization states in exfoliated Nb$_3$Cl$_8$ flakes and resolve a pronounced odd-even oscillation of the surface electrostatic potential. Beyond this parity-locked antiferroelectric order, we further identify intralayer polar domains in which local atomic reconstructions of the breathing kagome network reverse the out-of-plane dipole of the surface layer, producing ferroelectric-like stacking configurations. By interfacing monolayer MoSe$_2$ with Nb$_3$Cl$_8$, we demonstrate that these surface-polarization textures effectively modulate adjacent excitonic emission through domain-dependent interfacial band alignment and charge transfer. Our findings establish Nb$_3$Cl$_8$ as an intrinsic layer-polarized van der Waals platform and show that layer parity provides powerful structural degree of freedom for programming excitonic and optoelectronic responses at van der Waals interfaces.

\end{adjustwidth}
\end{abstract}
\date{\today}
\maketitle

\noindent

Layered van der Waals (vdW) materials provide an exceptional platform for engineering quantum and optoelectronic phenomena, as their atomically sharp interfaces can be integrated without the stringent lattice-matching constraints inherent to conventional heteroepitaxy \cite{geim2013van,novoselov20162d,ubrig2020design}.
In particular, polar and ferroelectric vdW materials are attractive building blocks for nanoscale devices, owing to their ability to modify the local electrostatic landscape, tune interfacial band alignment, and regulate charge transfer across adjacent layers\cite{yasuda2021stacking,kim2024electrostatic,chen2021ferroelectric,lin2020ferroelectric}.
Compared with externally imposed electrostatic gates or chemically engineered interfaces, an intrinsic polar surface offers a structurally defined and potentially nonvolatile means of controlling interfacial coupling. Identifying vdW crystals in which the surface polarization is determined by a simple and robust structural parameter is therefore important for developing programmable heterostructures.

\begin{figure*}[tbh]    
\includegraphics[width=2\columnwidth]{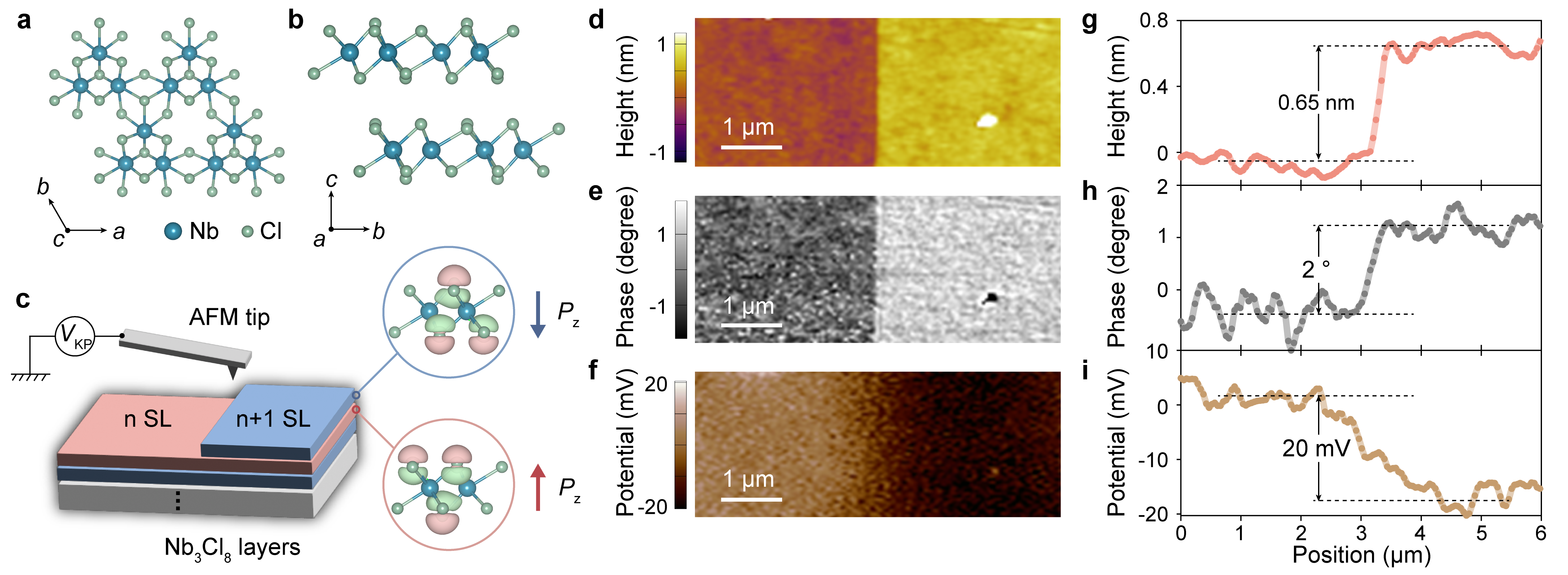}
\captionsetup{singlelinecheck=off, justification = RaggedRight}
\caption{\label{Figure1}\textbf{Layer-resolved surface polarization in Nb$_3$Cl$_8$.}
     \textbf{(a,b)} Top-view (a) and side-view (b) crystal structures of Nb$_3$Cl$_8$.
     \textbf{(c)} Schematic illustration of the KPFM measurement configuration. The red and blue slabs denote Nb$_3$Cl$_8$ monolayers with positive and negative out-of-plane polarization, respectively. The circular panels show the calculated charge-density differences for the corresponding polar layers.
     \textbf{(d--f)} AFM topography (d), AFM phase (e), and KPFM surface-potential (f) images of a Nb$_3$Cl$_8$ flake containing a single-layer step edge.
     \textbf{(g--i)} Line profiles extracted from the AFM topography (g), phase (h), and KPFM surface-potential (i) images, respectively.
}
\end{figure*}

A particularly appealing but still underexplored structural parameter is layer parity. In layered systems with alternating interlayer order, the net surface or bulk response can depend sensitively on whether the number of layers is odd or even. This effect is well established in layered antiferromagnets such as CrI$_3$\cite{huang2017layer}, CrSBr\cite{telford2020layered}, and MnBi$_2$Te$_4$\cite{yang2021odd}, where ferromagnetic order within each layer and antiferromagnetic coupling between neighboring layers produce an odd-even oscillation of the net magnetization. By contrast, layer-parity-dependent electric polarization has been much less explored, with only a few representative examples such as 2H-stacked $\alpha$-In$_2$Se$_3$, where the stacking-induced alternation of layer polarization gives rise to an odd-even dependence of the net ferroelectric response\cite{lv2021layer}. Extending this concept to other polar vdW crystals would provide a discrete, thickness-defined mechanism for controlling interfacial electrostatics.

Nb$_3$Cl$_8$ is a promising material in this context because its polar response originates from the intrinsic symmetry breaking of a breathing kagome lattice. In contrast to an ideal kagome network, the Nb sublattice in Nb$_3$Cl$_8$ undergoes trimerization, forming alternating small and large Nb triangles with inequivalent Nb--Nb lengths\cite{bolens2019topological,wang2023quantum} (Figure \ref{Figure1}a). This breathing distortion breaks inversion and mirror symmetries within each monolayer and provides a microscopic basis for an out-of-plane electric dipole. At the same time, Nb$_3$Cl$_8$ has attracted broad interest as a correlated layered material, owing to its cluster Mott insulating behavior and possible quantum spin-liquid physics\cite{gao2023discovery,liu2025direct,yang2025evidence,liu_possible_2024,fernando_strain-tunable_2026}. These features make Nb$_3$Cl$_8$ an unusual platform in which layer polarity, electron correlation, and vdW interfacial coupling can coexist.

Here, we combine atomic force microscopy (AFM), Kelvin probe force microscopy (KPFM), and first-principles calculations to investigate layer-parity-dependent surface polarization in exfoliated Nb$_3$Cl$_8$ and its interfacial effect on excitonic emission. We directly visualize an odd-even oscillation of the surface electrostatic potential, consistent with antiferroelectrically stacked polar layers, and further identify intralayer polar domains arising from local atomic reconstruction of the breathing kagome lattice. By integrating monolayer MoSe$_2$ with Nb$_3$Cl$_8$, we show that these polarization textures modulate interfacial band alignment and charge transfer, leading to spatially dependent excitonic responses. These results establish Nb$_3$Cl$_8$ as an intrinsic layer-polarized vdW platform and highlight layer parity as a structural degree of freedom for controlling optoelectronic phenomena at vdW interfaces.

\bigskip
\noindent \textbf{Layer-parity-dependent surface polarization in Nb$_3$Cl$_8$.} In the $\alpha$ phase of Nb$_3$Cl$_8$, neighboring polar layers are related by spatial inversion and stack in an AB sequence, resulting in antiparallel alignment of their out-of-plane dipoles. This antiferroelectric stacking cancels the internal dipoles in even-layer regions, leaving the topmost surface polarization determined by whether the total layer number is odd or even. Although bulk Nb$_3$Cl$_8$ is known to undergo a structural phase transition from the high-temperature $\alpha$ phase to the low-temperature $\beta$ phase\cite{haraguchi2017magnetic,sheckelton2017rearrangement,kim2023terahertz}, this transition can be strongly suppressed in thin exfoliated nanoflakes\cite{huang2026suppression}, preserving the layer-polarized $\alpha$ phase structure (Figure \ref{Figure1}b). Consequently, neighboring terraces separated by an odd number of layers are expected to exhibit opposite surface polarizations, whereas those differing by an even number of layers should retain the same surface polarization. Directly resolving this parity-dependent electrostatic landscape would provide a key signature of the intrinsic layer polarization in Nb$_3$Cl$_8$.

\begin{figure*}[tbh]    
\includegraphics[width=2\columnwidth]{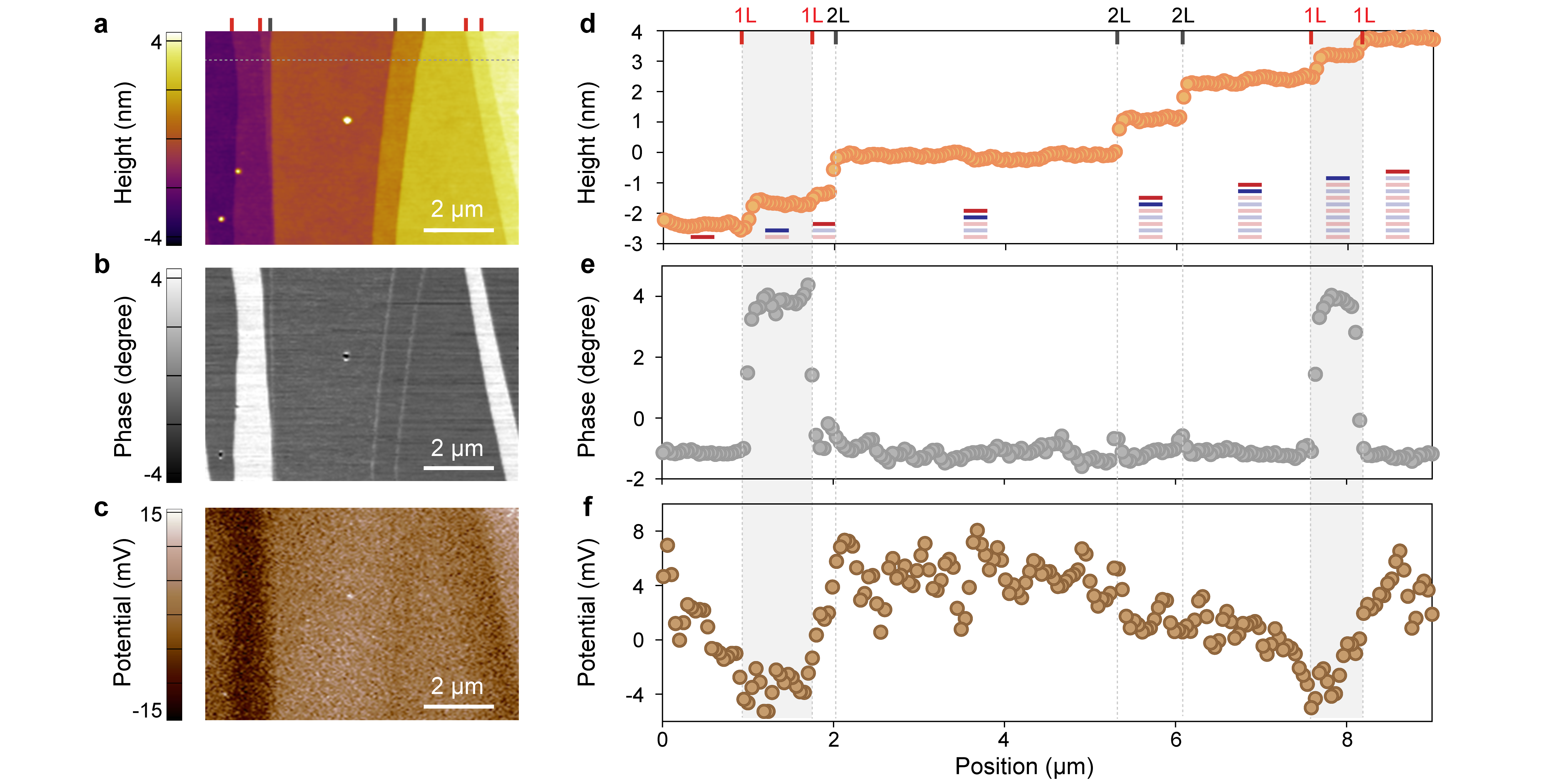}
\captionsetup{singlelinecheck=off, justification = RaggedRight}
\caption{\label{Figure2}\textbf{Visualization of layer-parity-dependent surface-polarization oscillations.}
     \textbf{(a--c)} AFM topography (a), AFM phase (b), and KPFM surface-potential (c) images of a Nb$_3$Cl$_8$ flake containing seven terraces separated by single- or double-layer steps.
     \textbf{(d--f)} Line profiles extracted from the AFM topography (d), AFM phase (e), and KPFM surface-potential (f) images, respectively. The vertical dashed gray lines mark the positions of the terrace boundaries. Insets in panel (d) schematically illustrate the out-of-plane polarization configuration of the Nb$_3$Cl$_8$ layers, where red and blue denote positive and negative $P_\mathrm{z}$, respectively.
}
\end{figure*}

All Nb$_3$Cl$_8$ flakes were prepared by mechanical exfoliation (see Methods), with thicknesses well below 20 nm (see Supplementary Information Fig. S1). To verify the existence of out-of-plane electric polarization ($P_\text{z}$) within individual Nb$_3$Cl$_8$ layers, we calculated the charge density difference of the distorted breathing kagome structure relative to an undistorted reference structure with equivalent Nb--Nb bond lengths (circular panels in Figure \ref{Figure1}c). The resulting asymmetric charge redistribution between the upper and lower Cl atomic planes within a single Nb$_3$Cl$_8$ layer indicates the emergence of a finite out-of-plane electric polarization. Combined with the inversion-related stacking of adjacent layers, this intrinsic layer dipole naturally gives rise to the antiferroelectric configuration and parity-dependent surface polarization illustrated in Figure \ref{Figure1}c.

To experimentally verify the layer-dependent electric polarization, we characterize a $\sim$15.9-nm-thick Nb$_3$Cl$_8$ nanoflake containing a step edge using AFM and KPFM (Figure \ref{Figure1}c), which allow simultaneous visualization of the surface topography and local electrostatic potential distribution (see Methods for details). Figures \ref{Figure1}d--f display the AFM topography, phase, and electrostatic potential maps acquired across the terrace. The step height of 0.65 nm extracted from the topographic profile confirms that the thickness difference between the two adjacent regions corresponds to a single Nb$_3$Cl$_8$ layer\cite{sun2022observation}. Across the same step edge, both the AFM phase and KPFM surface-potential maps exhibit clear contrasts, providing evidence for layer-dependent surface polarization in Nb$_3$Cl$_8$. The measured potential difference of approximately 20 mV is consistent with opposite out-of-plane polarization orientations of the topmost Nb$_3$Cl$_8$ layers on the two adjacent terraces, allowing the relative surface-polarization direction to be identified. The AFM phase contrast is sharper than the corresponding KPFM potential contrast and is therefore useful for locating abrupt boundaries. However, tapping-mode AFM phase shifts generally reflect the full tip-sample interaction including both conservative forces (e.g., elastic, electrostatic and magnetic contributions) and dissipative processes (e.g. friction, viscoelasticity and adhesion hysteresis). We therefore interpret the phase images in conjunction with KPFM, rather than assigning the phase contrast solely to electrostatic polarization\cite{garcia1998phase,melcher2009origins}.

\begin{figure*}[tbh]    
\includegraphics[width=2\columnwidth]{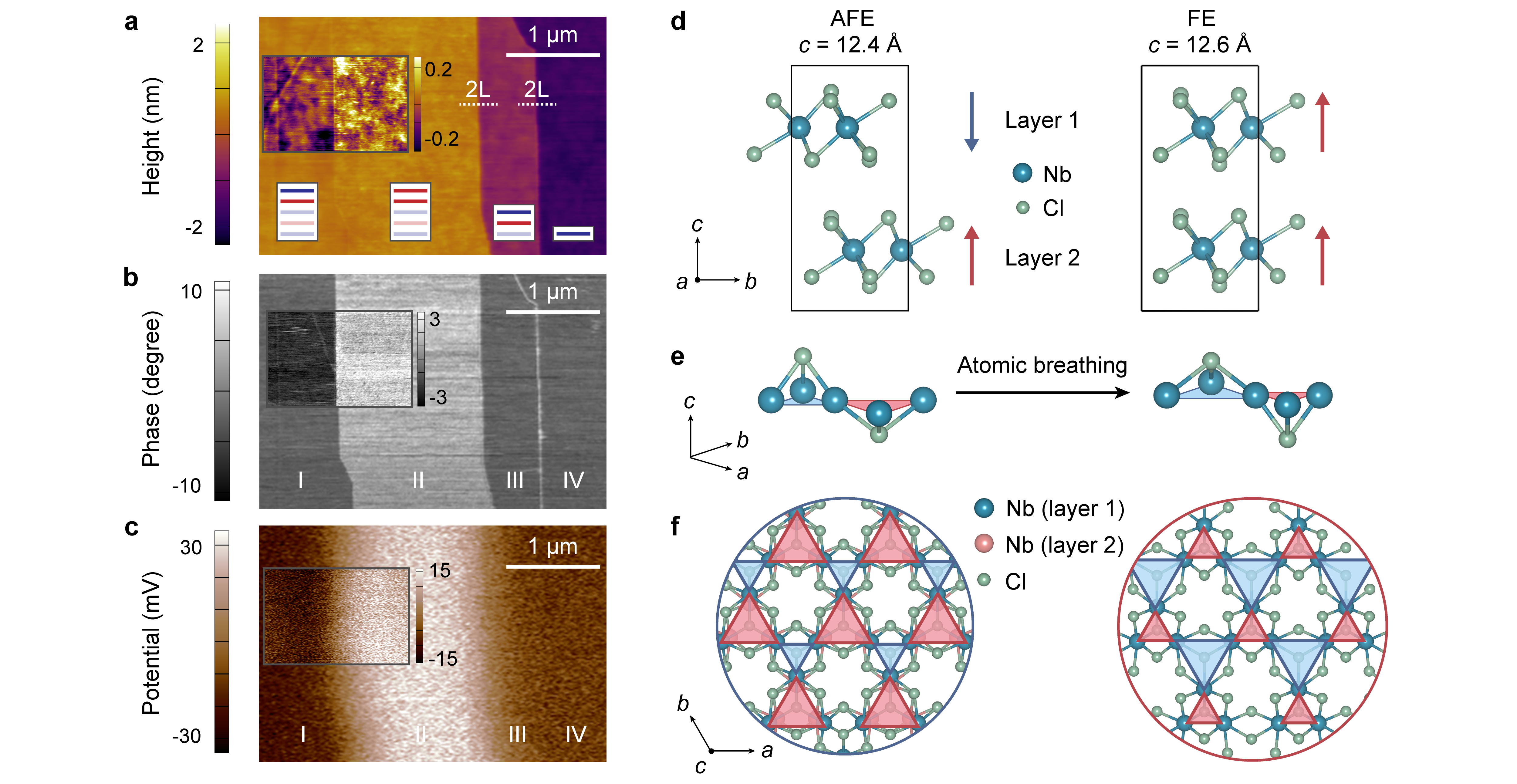}
\captionsetup{singlelinecheck=off, justification = RaggedRight}
\caption{\label{Figure3}\textbf{Intralayer polar reconstruction in Nb$_3$Cl$_8$.}
     \textbf{(a--c)} AFM topography (a), AFM phase (b), and KPFM surface-potential (c) images of a Nb$_3$Cl$_8$ flake showing a layer-parity-independent polar region and two bilayer terraces. The inset panels show zoom-in scans of the boxed regions with adjusted color scales to highlight the local contrast. Schematics in panel (a) depict the corresponding out-of-plane polarization configurations of the Nb$_3$Cl$_8$ layers, where red and blue represent positive and negative $P_\mathrm{z}$, respectively.
     \textbf{(d)} Optimized bilayer crystal structures for the antiferroelectric-like (AFE) and ferroelectric-like (FE) stacking configurations of Nb$_3$Cl$_8$. Red and blue arrows indicate the polarization orientations of individual layers.
     \textbf{(e)} Schematic illustration of intralayer polarization switching associated with breathing distortions of the Nb-trimer kagome network.
     \textbf{(f)} Top-view structural models corresponding to the breathing reconstruction illustrated in (e).
}
\end{figure*}

We further demonstrate the layer-parity-dependent polarization oscillation by examining an additional Nb$_3$Cl$_8$ flake with multiple step edges. As shown in the AFM topography image (Figure \ref{Figure2}a), the sample surface consists of seven distinct terraces. The height profile extracted from the topography (dashed gray line in Figure \ref{Figure2}a) verifies that the step heights across the terraces correspond to either single-layer (1L) or double-layer (2L) thickness differences. The corresponding AFM phase (Figure \ref{Figure2}b) and electrostatic potential (Figure \ref{Figure2}c) maps provide complementary information on the surface-polarization state. Both measurements exhibit a highly correlated, parity-dependent behavior. Specifically, crossing step edges associated with an odd-layer-number difference (e.g., 1L) gives rise to pronounced and reproducible contrasts in both the AFM phase ($\sim$5$^\circ$) and surface potential ($\sim$10 mV). In contrast, across step edges corresponding to an even-layer-number difference (e.g., 2L), the phase and surface-potential signals remain essentially unchanged. These observations establish a robust odd-even oscillation of the surface electrostatic landscape, which has been consistently reproduced in multiple independent samples (see Supplementary Information Fig. S2).

\bigskip
\noindent \textbf{Intralayer polar reconstruction associated with kagome breathing distortions.} In addition to the ideal antiferroelectric order defined by layer parity, the weak interlayer coupling in the staggered Nb-trimer stacking of $\alpha$-Nb$_3$Cl$_8$ may allow local polar reconstruction within individual layers. Figure \ref{Figure3} presents a representative case in which the observed polarization configuration cannot be fully understood within the framework of the parity-defined antiferroelectric ground state. The AFM topography image (Figure \ref{Figure3}a) reveals two bilayer terraces, with the thickness differences further verified by the height profiles shown in Supplementary Information Fig. S3. According to the layer-parity-dependent surface-polarization rule established above, step edges associated with an even-number layer difference should preserve the same surface-polarization state and therefore should not produce pronounced changes in either the AFM phase or KPFM surface-potential signals. However, a pronounced enhancement in both signals is observed in the central region, denoted as domain II in Figures \ref{Figure3}b,c. Notably, domain II has no layer difference from the region on its left side (a 0L boundary), and differs from the region on its right side by two layers (a 2L step). Neither boundary should lead to a polarity reversal according to the layer-parity rule. The clear phase and surface-potential contrasts associated with domain II are therefore inconsistent with a purely layer-parity-driven origin, suggesting that local intralayer atomic reconstruction modifies the intrinsic surface-polarization configuration of Nb$_3$Cl$_8$.

Using the layer-parity-dependent surface-polarization behavior established above as a reference, the relative polarization direction of the surface Nb$_3$Cl$_8$ layers can be assigned from the KPFM electrostatic potential contrast. Domains I, III, and IV correspond to a negative out-of-plane polarization state ($P_\text{z}<0$), whereas domain II exhibits the opposite, positive polarization state ($P_\text{z}>0$). Insets of Figures \ref{Figure3}a--c show zoom-in scans of the left boundary between domains I and II, where a clear reversal of the surface-polarization contrast is observed. Importantly, this boundary does not involve any change in the Nb$_3$Cl$_8$ layer number, indicating that the polarization reversal is not driven by layer parity. Nevertheless, the corresponding topography map (inset of Figure \ref{Figure3}a) reveals a subtle but reproducible height variation across the boundary. Statistical analysis of averaged height profiles extracted from ten independent linecuts shows that domain II is approximately 0.5 $\text{\AA}$ higher than domain I (see Supplementary Information Fig. S4). This small yet systematic height offset suggests that the polarization reversal is accompanied by a local structural reconstruction, likely involving variations in interlayer spacing or stacking registry.

\begin{figure*}[tbh]    
\includegraphics[width=1.7\columnwidth]{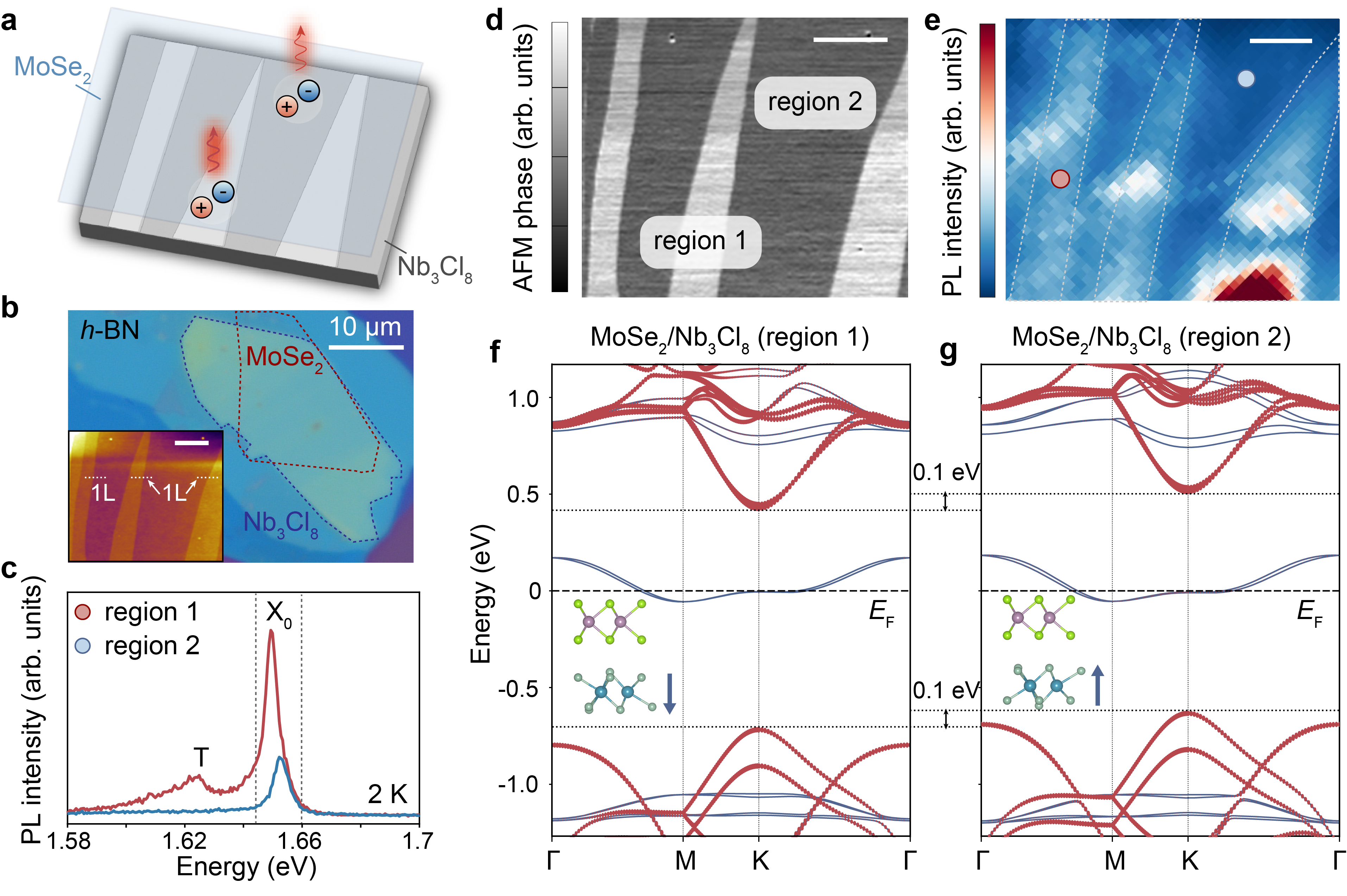}
\captionsetup{singlelinecheck=off, justification = RaggedRight}
\caption{\label{Figure4}\textbf{Layer-parity-controlled excitonic emission at the MoSe$_2$/Nb$_3$Cl$_8$ interface.}
     \textbf{(a)} Schematic illustration of MoSe$_2$ excitonic emission modulated by the layer-parity-defined surface polarization of the underlying Nb$_3$Cl$_8$.
     \textbf{(b)} Optical microscopy image of the MoSe$_2$/Nb$_3$Cl$_8$ heterostructure encapsulated by $\textit{h}$-BN flakes. Inset: AFM topography image of the Nb$_3$Cl$_8$ substrate. Scale bar: 2 $\mu$m.
     \textbf{(c)} PL spectra of the MoSe$_2$/Nb$_3$Cl$_8$ heterostructure acquired from region 1 and region 2 at 2 K.
     \textbf{(d)} AFM phase image of the Nb$_3$Cl$_8$ substrate, revealing ribbon-like regions separated from the surrounding areas by monolayer step edges. Scale bar: 2 $\mu$m.
     \textbf{(e)} Spatial map of the integrated PL intensity of the heterostructure over the energy window indicated by the gray dashed lines in panel (c). The red and blue circles denote the positions where the PL spectra in panel (c) were acquired. Boundaries of the ribbon-like regions are highlighted by white dashed lines. Scale bar: 2 $\mu$m.
     \textbf{(f,g)} DFT-calculated electronic band structures of the heterostructure with the region 1 (f) and region 2 (g) interfacial configurations, respectively. Insets show the corresponding atomic structures.
}
\end{figure*}

On the basis of these observations, we propose that the top two Nb$_3$Cl$_8$ layers in domain II locally adopt a ferroelectric-like stacking configuration, in which their out-of-plane polarizations are aligned in the same upward direction (Figure \ref{Figure3}d). First-principles calculations support this interpretation by yielding a stable ferroelectric-stacked bilayer structure with a $c$-axis lattice constant approximately 0.2 $\text{\AA}$ larger than that of the pristine antiferroelectric-stacked $\alpha$-Nb$_3$Cl$_8$ bilayer. This calculated lattice expansion is qualitatively consistent with the experimentally observed AFM height increase in domain II. The quantitative difference between the calculated and measured height changes may arise from several factors, including surface relaxation, tip-sample convolution, and local topographic fluctuations.

We further attribute the intralayer polarization switching between domains I and II to local atomic reconstruction associated with the breathing distortion of Nb atoms (Figures \ref{Figure3}e,f). In this reconstruction, the initially small Nb triangles expand into larger ones (blue), whereas the initially large Nb triangles contract into smaller ones (red), thereby reversing the local layer dipole orientation. Such a reconstruction may be facilitated by localized mechanical strain or stacking perturbations introduced during the exfoliation process.

\bigskip
\noindent \textbf{Layer-parity-controlled excitonic emission at MoSe$_2$/Nb$_3$Cl$_8$ interfaces.} The layer-parity-dependent surface polarization in Nb$_3$Cl$_8$ provides a promising platform for realizing spatially programmable interfacial functionalities. To demonstrate this capability, we fabricated a vdW heterostructure by transferring a monolayer MoSe$_2$ onto a Nb$_3$Cl$_8$ flake containing multiple layer-defined terraces, and explored how the local surface-polarization state modulates excitonic emission (Figures \ref{Figure4}a,b). As revealed by the AFM topography (inset of Figure \ref{Figure4}b), the Nb$_3$Cl$_8$ substrate contains three characteristic ribbon-like regions that are exactly one layer thicker than the surrounding areas. These one-layer-thicker ribbons show pronounced contrast in both the AFM phase and electrostatic potential maps (see Figure \ref{Figure4}d and Supplementary Information Fig. S5), indicating a surface-polarization orientation opposite to that of the adjacent thinner regions, consistent with the layer-parity-dependent polarization order.

Photoluminescence (PL) measurements were performed at 2 K to characterize the excitonic emission of the heterostructure. The neutral exciton (X$_0$) of monolayer MoSe$_2$ is located at approximately 1.65 eV\cite{ross2013electrical}. Figure \ref{Figure4}e presents the spatial map of the integrated PL intensity within the neutral exciton energy range (1.645--1.655 eV), which exhibits a pronounced ribbon-like contrast closely matching the AFM phase map. This strong spatial correlation indicates that Nb$_3$Cl$_8$ surface regions with opposite layer-parity-defined polarization states can effectively modulate the excitonic emission of the adjacent MoSe$_2$ monolayer. Representative PL spectra acquired from MoSe$_2$ on negatively and positively polarized Nb$_3$Cl$_8$ regions (denoted as regions 1 and 2, respectively) are shown in Figure \ref{Figure4}c, with schematic illustrations of the corresponding interfacial configurations shown in the insets of Figures \ref{Figure4}f,g. The PL spectrum acquired from region 1 exhibits a neutral exciton emission together with a discernible low-energy trion feature (T) at around 1.625 eV, indicating the presence of residual electron doping in this MoSe$_2$ region. In contrast, region 2 shows a single, well-defined neutral exciton peak with strongly suppressed trion emission, suggesting that MoSe$_2$ becomes more charge neutral in this interfacial configuration. This behavior demonstrates that the layer-parity-defined surface polarization of Nb$_3$Cl$_8$ gives rise to region-dependent interfacial charge transfer and carrier doping in the MoSe$_2$ monolayer.

To elucidate the mechanism underlying the layer-parity-dependent excitonic modulation, density functional theory (DFT) calculations were conducted for MoSe$_2$/Nb$_3$Cl$_8$ heterostructures with interfacial configurations corresponding to region 1 and region 2, as illustrated in Figures \ref{Figure4}f,g. Although correlation effects are not explicitly included in these calculations, they provide insight into how the electronic structure evolves across the two configurations. Comparison of the calculated band structures reveals that the overall band dispersions of both MoSe$_2$ and Nb$_3$Cl$_8$ remain nearly unchanged between the two configurations. Instead, the primary difference lies in the relative band alignment between the two materials. 

In the region 1 configuration, the MoSe$_2$ bands are shifted downward by approximately 0.1 eV relative to those in the region 2 configuration, with respect to the Nb$_3$Cl$_8$ bands. This band alignment places the Fermi level closer to the MoSe$_2$ conduction-band edge in region 1, favoring a more electron-doped MoSe$_2$ layer, consistent with the pronounced trion emission observed experimentally. By contrast, the band alignment in region 2 favors more efficient electron transfer from MoSe$_2$ to Nb$_3$Cl$_8$, thereby reducing the excess electron density in MoSe$_2$ and making its PL response more dominated by neutral excitons. This interpretation is further supported by the PL spectrum of bare monolayer MoSe$_2$, which already exhibits a strong trion feature (see Supplementary Data Fig. S6), indicating substantial intrinsic n-type doping before forming the heterostructure. The suppression of the trion feature in region 2 therefore suggests that additional electrons are transferred from MoSe$_2$ into Nb$_3$Cl$_8$, driving MoSe$_2$ toward a more charge-neutral excitonic state.

\bigskip
\noindent 
\textbf{Conclusion}\\
In summary, our work demonstrates that exfoliated Nb$_3$Cl$_8$ hosts robust layer-parity-dependent surface polarization originating from the antiferroelectric stacking of intrinsically polar breathing-kagome layers. Combined AFM and KPFM measurements reveal pronounced odd-even oscillations of the surface electrostatic landscape across multilayer terraces, establishing layer parity as a discrete structural degree of freedom for controlling polar surface states in Nb$_3$Cl$_8$. In addition to this intrinsic parity-defined order, we identify local intralayer polar reconstruction associated with Nb-trimer breathing distortions and modified stacking registry, giving rise to polarization textures that are not dictated solely by layer number. By integrating monolayer MoSe$_2$ with Nb$_3$Cl$_8$, we further demonstrate that layer-parity-defined surface polarization modulates interfacial excitonic emission through polarization-dependent band alignment and charge transfer. Notably, this layer parity can also alter the transport behavior of heterostructures, such as the monolayer graphene/Nb$_3$Cl$_8$ heterostructure. 
These findings establish Nb$_3$Cl$_8$ as an intrinsic layer-polarized vdW platform for programmable interface engineering and open opportunities for controlling excitonic and optoelectronic responses through discrete polar degrees of freedom in vdW heterostructures.

\bigskip
\noindent 
\textbf{Methods}\\
\noindent \textbf{Crystal synthesis.} 
Bulk MoSe$_2$ single crystals were synthesized via the chemical vapor transport (CVT) method. Initially, polycrystalline MoSe$_2$ was prepared by reacting a stoichiometric mixture of Mo and Se powders (1:2 molar ratio) at 700 $^{\circ}$C for 120 h, following a 48 h ramping period. The resulting precursor was carefully ground into powder. Subsequently, 1000 mg of the synthesized MoSe$_2$ powder, along with Se ($14\text{ mg/mL}$) acting as the self-transport agent, was thoroughly blended and vacuum-sealed inside a quartz tube. The tube was then placed in a dual-zone furnace and subjected to a temperature gradient from 1020 $^{\circ}$C (source) to 960 $^{\circ}$C (growth) for 216 h.

Single crystals of Nb$_3$Cl$_8$ were grown via a PbCl$_2$-flux-assisted technique. First, the precursor powder was obtained through a solid-state reaction of high-purity Nb (Alfa Aesar, 99.99\%) and NbCl$_5$ (Alfa Aesar, 99.9\%) in a 7:8 molar ratio. The reagents were thoroughly mixed, inside an evacuated quartz tube, and annealed at 700 $^\circ$C for 48 h. The resulting intermediate was then combined with an appropriate amount of $\text{PbCl}_2$ flux and re-encapsulated under equivalent vacuum conditions. This mixture was heated to 750 $^\circ$C over 20 h, maintained at this temperature for 300 h, and subsequently cooled to 500 $^\circ$C over a 100 h period. After naturally equilibrating to ambient temperature, the $\text{Nb}_3\text{Cl}_8$ crystals were retrieved by removing the residual flux with hot deionized water.\\

\noindent \textbf{Sample preparation.} 
Two-dimensional Nb$_3$Cl$_8$, monolayer MoSe$_2$, and \textit{h}-BN nanoflakes were mechanically exfoliated from their bulk counterparts onto Si/SiO$_2$ substrates under ambient conditions. Monolayer MoSe$_2$ flakes were initially identified by optical contrast and subsequently verified through room-temperature PL spectroscopy. Using a layer-by-layer dry transfer method, the MoSe$_2$/Nb$_3$Cl$_8$ heterostructure was assembled with a poly (bisphenol A carbonate) film supported on a dome-shaped polydimethylsiloxane (PDMS) stamp. \\

\noindent \textbf{KPFM characterizations.} 
KPFM measurements were performed using an Asylum Research Cypher S atomic force microscope. Conductive ASYELEC.01-R2 probes were used for the imaging, with a  mechanical resonance frequency of $\sim$75 kHz and a force constant of $\sim$2.8 N/m. The measurements were performed in a dual-pass Nap mode. In the first pass, the topography and phase of the sample surface was recorded in standard tapping mode. In the subsequent second pass (Nap line), the probe was lifted to a well-defined constant height (typically 40 nm) above the surface to detect long-range electrostatic forces while eliminating short-range van der Waals interactions. During this second pass, the cantilever was electrically driven by an electrical voltage that consists of DC and AC components. The total voltage difference $V$ between the tip and the 
sample is expressed as\cite{nonnenmacher1991kelvin,melitz2011kelvin}:
\begin{equation}
    V = (V_\text{DC} - V_\text{sp}) + V_\text{AC}\sin(\omega t) 
\end{equation}
where $V_{\rm{DC}}$ is the applied direct current bias and $V_{\rm{sp}}$ is the local surface potential of the sample. This voltage induces an electrostatic force given by:
\begin{equation}
\begin{split}
F = \frac{1}{2} \frac{\partial C}{\partial z}
\Big\{ &[(V_\text{DC} - V_\text{sp})^2+\frac{1}{2}V_\text{AC}^2] \\ 
&+2[(V_\text{DC}-V_\text{sp})V_\text{AC}\sin(\omega t)] \\
&-[\frac{1}{2}V_\text{AC}^2\cos(2\omega t)]
\Big\}
\end{split}
\end{equation}
To extract the surface potential, a lock-in amplifier feedback loop was utilized to adjust the $V_{\rm{DC}}$ bias until the cantilever's vibration amplitude at the first harmonic frequency ($\omega$) was minimized to zero. Under this condition, the applied DC bias satisfies $V_\text{DC} = V_\text{sp}$. Spatial mapping of this feedback-controlled $V_\text{DC}$ yielded the local surface potential images.
\\

\noindent \textbf{Optical measurements.} 
The optical measurements were carried out in a closed-cycle helium cryostat (attoDRY2100) with a base temperature of 1.6 K. PL spectroscopy was performed in a backscattering configuration, using a HeNe laser (1.96 eV) as the excitation source. The beam was focused onto the sample through a high numerical aperture objective ($\text{NA} = 0.82$) , yielding a focal spot diameter of approximately 1 $\mu$m. The emitted PL signal was collected by a spectrometer (SpectraPro HRS-500S), dispersed using a 150 grooves/mm grating, and detected by a liquid-nitrogen-cooled charge-coupled device (PyLoN:400). Spatial PL mapping was achieved by scanning the sample with a high-precision $x-y$ piezoelectric sample stage. \\

\noindent \textbf{First-principles calculations.}
The first-principles calculations based on density functional theory were performed using the projected-augmented-wave (PAW) pseudopotential and the generalized gradient approximation (GGA) of Perdew-Burke-Ernzerhof (PBE) functional, as implemented in the Vienna ab initio simulation package (VASP)\cite{kresse1996efficient,kresse1996efficiency}. The plane-wave energy cutoff was set to 500 eV, and the convergence criterion for the total energy was set to 10$^{-8}$ eV. All geometries were optimized using the conjugate gradient method, until none of the residual Hellmann-Feynman forces exceeds 10$^{-3}$ eV/$\text{\AA}$. The spin-orbit coupling (SOC) effect was taken into account in all calculations. To avoid fictitious Coulomb interaction between periodically repeated images for 2D systems, a large vacuum slab of 10 $\text{\AA}$ thickness was employed on each side of the atomic layers. Both Nb$_3$Cl$_8$ and MoSe$_2$ crystallize in hexagonal lattice, and with 9$\times$9$\times$1 and 18$\times$18$\times$1 Monkhorst-Pack k-point meshes adopted respectively, the equilibrium lattice constants after monolayer relaxation were 6.77 $\text{\AA}$ and 3.32 $\text{\AA}$, respectively. The undistorted phase of Nb$_3$Cl$_8$ was generated by fixing the lattice parameters while adopting a nonpolar atomic configuration. The MoSe$_2$/Nb$_3$Cl$_8$ heterostructures were built using a 2$\times$2 MoSe$_2$ supercell and a Nb$_3$Cl$_8$ unit cell, resulting in a lattice mismatch of only 1.92\%. Different interfacial configurations were considered. For the heterostructures, only the atomic positions were allowed to relax and the DFT-D2 method was used to account for the van der Waals interactions between the layers.\\

\bigskip
\noindent\textbf{Data availability}\\
\noindent
All relevant data are available in the main text, Supporting Information, or upon request to the authors.\\

\bigskip
\noindent\textbf{Acknowledgment}\\
\noindent
This work was supported by the National Natural Science Foundation of China (No. 12425402) and the National Key R\&D Program of China (No. 2025YFA1411002 and No. 2022YFA1203902). K.W. and T.T. acknowledge support from the JSPSKAKENHI (No. 21H05233 and No. 23H02052) and World Premier International Research Center Initiative (WPI), MEXT, Japan. \\

\bigskip
\noindent\textbf{Author contributions}\\
\noindent
Y.Y. and X.H. conceived the project and designed the experiments. X.H. and H.X. prepared the Nb$_3$Cl$_8$ samples and performed the KPFM characterizations, with support from Y.G.. X.H. and Y.Z. performed the MoSe$_2$/Nb$_3$Cl$_8$ optical measurements. J.H. and T.Y. carried out the first-principles calculations. Z.C. grew the MoSe$_2$ single crystal. Z. M. and Y.S. synthesized the Nb$_3$Cl$_8$ bulk crystals. K.W. and T.T. grew the \textit{h}-BN single crystals. X.H. and Y.Y. wrote the manuscript. All authors discussed the results and contributed to the manuscript. \\

\bigskip
\noindent\textbf{Competing interests}\\
\noindent
The authors declare no competing interests.\\

\bigskip
\noindent\textbf{Additional information}\\
\noindent
\textbf{Supporting information} The online version contains Supporting material available at URL.\\

\normalem
\bibliographystyle{naturemag}
\bibliography{ref}

@article{geim2013van,
  title={Van der Waals heterostructures},
  author={Geim, Andre K and Grigorieva, Irina V},
  journal={Nature},
  volume={499},
  number={7459},
  pages={419--425},
  year={2013},
  publisher={Nature Publishing Group UK London}
}

@article{novoselov20162d,
  title={2{D} materials and van der Waals heterostructures},
  author={Novoselov, K Sꎬ and Mishchenko, Artem and Carvalho, Alexandra and Castro Neto, AH},
  journal={Science},
  volume={353},
  number={6298},
  pages={aac9439},
  year={2016},
  publisher={American Association for the Advancement of Science}
}

@article{ubrig2020design,
  title={Design of van der Waals interfaces for broad-spectrum optoelectronics},
  author={Ubrig, Nicolas and Ponomarev, Evgeniy and Zultak, Johanna and Domaretskiy, Daniil and Z{\'o}lyomi, Viktor and Terry, Daniel and Howarth, James and Guti{\'e}rrez-Lezama, Ignacio and Zhukov, Alexander and Kudrynskyi, Zakhar R and others},
  journal={Nature Materials},
  volume={19},
  number={3},
  pages={299--304},
  year={2020},
  publisher={Nature Publishing Group UK London}
}

@article{yasuda2021stacking,
  title={Stacking-engineered ferroelectricity in bilayer boron nitride},
  author={Yasuda, Kenji and Wang, Xirui and Watanabe, Kenji and Taniguchi, Takashi and Jarillo-Herrero, Pablo},
  journal={Science},
  volume={372},
  number={6549},
  pages={1458--1462},
  year={2021},
  publisher={American Association for the Advancement of Science}
}

@article{kim2024electrostatic,
  title={Electrostatic moir{\'e} potential from twisted hexagonal boron nitride layers},
  author={Kim, Dong Seob and Dominguez, Roy C. and Mayorga-Luna, Rigo and Ye, Dingyi and Embley, Jacob and Tan, Tixuan and Ni, Yue and Liu, Zhida and Ford, Mitchell and Gao, Frank Y. and Yang, Li and Li, Xiaoqin Elaine and Miyahara, Yoichi},
  journal={Nature Materials},
  volume={23},
  number={1},
  pages={65--70},
  year={2024},
  publisher={Nature Publishing Group},
  doi={10.1038/s41563-023-01637-7}
}

@article{chen2021ferroelectric,
  title={Ferroelectric-tuned van der Waals heterojunction with band alignment evolution},
  author={Chen, Yan and Wang, Xudong and Huang, Le and Wang, Xiaoting and Jiang, Wei and Wang, Zhen and Wang, Peng and Wu, Binmin and Lin, Tie and Shen, Hong and others},
  journal={Nature Communications},
  volume={12},
  number={1},
  pages={4030},
  year={2021},
  publisher={Nature Publishing Group}
}

@article{lin2020ferroelectric,
  title={Ferroelectric-field accelerated charge transfer in 2D CuInP2S6 heterostructure for enhanced photocatalytic H2 evolution},
  author={Lin, Bo and Chaturvedi, Apoorva and Di, Jun and You, Lu and Lai, Chen and Duan, Ruihuan and Zhou, Jiadong and Xu, Baorong and Chen, Zihao and Song, Pin and others},
  journal={Nano Energy},
  volume={76},
  pages={104972},
  year={2020},
  publisher={Elsevier}
}

@article{huang2017layer,
  title={Layer-dependent ferromagnetism in a van der Waals crystal down to the monolayer limit},
  author={Huang, Bevin and Clark, Genevieve and Navarro-Moratalla, Efr{\'e}n and Klein, Dahlia R and Cheng, Ran and Seyler, Kyle L and Zhong, Ding and Schmidgall, Emma and McGuire, Michael A and Cobden, David H and others},
  journal={Nature},
  volume={546},
  number={7657},
  pages={270--273},
  year={2017},
  publisher={Nature Publishing Group UK London}
}

@article{telford2020layered,
  title={Layered antiferromagnetism induces large negative magnetoresistance in the van der Waals semiconductor CrSBr},
  author={Telford, Evan J and Dismukes, Avalon H and Lee, Kihong and Cheng, Minghao and Wieteska, Andrew and Bartholomew, Amymarie K and Chen, Yu-Sheng and Xu, Xiaodong and Pasupathy, Abhay N and Zhu, Xiaoyang and others},
  journal={Advanced Materials},
  volume={32},
  number={37},
  pages={2003240},
  year={2020},
  publisher={Wiley Online Library}
}

@article{yang2021odd,
  title={Odd-even layer-number effect and layer-dependent magnetic phase diagrams in {MnBi$_2$Te$_4$}},
  author={Yang, Shiqi and Xu, Xiaolong and Zhu, Yaozheng and Niu, Ruirui and Xu, Chunqiang and Peng, Yuxuan and Cheng, Xing and Jia, Xionghui and Huang, Yuan and Xu, Xiaofeng and others},
  journal={Physical Review X},
  volume={11},
  number={1},
  pages={011003},
  year={2021},
  publisher={APS}
}

@article{lv2021layer,
  title={Layer-dependent ferroelectricity in 2H-stacked few-layer $\alpha$-{In$_2$Se$_3$}},
  author={Lv, Baohua and Yan, Zhi and Xue, Wuhong and Yang, Ruilong and Li, Jiayi and Ci, Wenjuan and Pang, Ruixue and Zhou, Peng and Liu, Gang and Liu, Zhongyuan and others},
  journal={Materials Horizons},
  volume={8},
  number={5},
  pages={1472--1480},
  year={2021},
  publisher={Royal Society of Chemistry}
}

@article{bolens2019topological,
  title={Topological states on the breathing kagome lattice},
  author={Bolens, Adrien and Nagaosa, Naoto},
  journal={Physical Review B},
  volume={99},
  number={16},
  pages={165141},
  year={2019},
  publisher={APS}
}

@article{wang2023quantum,
  title={Quantum states and intertwining phases in kagome materials},
  author={Wang, Yaojia and Wu, Heng and McCandless, Gregory T and Chan, Julia Y and Ali, Mazhar N},
  journal={Nature Reviews Physics},
  volume={5},
  number={11},
  pages={635--658},
  year={2023},
  publisher={Nature Publishing Group UK London}
}

@article{gao2023discovery,
  title={Discovery of a single-band Mott insulator in a van der Waals flat-band compound},
  author={Gao, Shunye and Zhang, Shuai and Wang, Cuixiang and Yan, Shaohua and Han, Xin and Ji, Xuecong and Tao, Wei and Liu, Jingtong and Wang, Tiantian and Yuan, Shuaikang and others},
  journal={Physical Review X},
  volume={13},
  number={4},
  pages={041049},
  year={2023},
  publisher={APS}
}

@article{liu2025direct,
  title={Direct evidence of intrinsic Mott state and its layer-parity oscillation in a breathing kagome crystal down to monolayer},
  author={Liu, Huanyu and Li, Wenhui and Zhou, Zishu and Qu, Hongbin and Zhang, Jiaqi and Hu, Weixiong and Wen, Chenhaoping and Wang, Ning and Deng, Hao and Li, Gang and others},
  journal={Physical Review Letters},
  volume={135},
  number={7},
  pages={076503},
  year={2025},
  publisher={APS}
}

@article{yang2025evidence,
  title={Evidence of Mott insulator with thermally induced melting behavior in kagome compound $\text{Nb}_{3}\text{Cl}_8$},
  author={Yang, Qiu and Wu, Min and Duan, Jingyi and Ma, Zhijie and Li, Lingxiao and Huo, Zihao and Zhang, Zaizhe and Watanabe, Kenji and Taniguchi, Takashi and Zhao, Xiaoxu and others},
  journal={National Science Review},
  volume={12},
  number={12},
  pages={nwaf464},
  year={2025},
  publisher={Oxford University Press}
}

@article{fernando_strain-tunable_2026,
	title = {Strain-tunable magnetic correlations in spin liquid candidate $\text{Nb}_{3}\text{Cl}_8$},
	volume = {13},
	issn = {2053-1583},
	doi = {10.1088/2053-1583/ae4b4e},
	language = {en},
	number = {2},
	urldate = {2026-03-10},
	journal = {2D Materials},
	author = {Fernando, Tharindu and Cao, Ting},
	month = mar,
	year = {2026},
	pages = {025015},
}

@article{liu_possible_2024,
	title = {Possible quantum-spin-liquid state in van der {Waals} cluster magnet $\text{Nb}_{3}\text{Cl}_8$},
	volume = {36},
	issn = {0953-8984},
	doi = {10.1088/1361-648X/ad1a5c},
	language = {en},
	number = {15},
	urldate = {2024-03-19},
	journal = {Journal of Physics: Condensed Matter},
	author = {Liu, Bo and Zhang, Yongchao and Han, Xin and Sun, Jianping and Zhou, Honglin and Li, Chunhong and Cheng, Jinguang and Yan, Shaohua and Lei, Hechang and Shi, Youguo and Yang, Huaixin and Li, Shiliang},
	month = jan,
	year = {2024},
	pages = {155602},
}

@article{haraguchi2017magnetic,
  title={Magnetic-nonmagnetic phase transition with interlayer charge disproportionation of {Nb$_3$} trimers in the cluster compound Nb$_3$Cl$_8$},
  author={Haraguchi, Yuya and Michioka, Chishiro and Ishikawa, Manabu and Nakano, Yoshiaki and Yamochi, Hideki and Ueda, Hiroaki and Yoshimura, Kazuyoshi},
  journal={Inorganic Chemistry},
  volume={56},
  number={6},
  pages={3483--3488},
  year={2017},
  publisher={ACS Publications}
}

@article{sheckelton2017rearrangement,
  title={Rearrangement of van der Waals stacking and formation of a singlet state at {$T = 90 \text{K}$} in a cluster magnet},
  author={Sheckelton, John P and Plumb, Kemp W and Trump, Benjamin A and Broholm, Collin L and McQueen, Tyrel M},
  journal={Inorganic Chemistry Frontiers},
  volume={4},
  number={3},
  pages={481--490},
  year={2017},
  publisher={Royal Society of Chemistry}
}

@article{kim2023terahertz,
  title={Terahertz spectroscopy and dft analysis of phonon dynamics of the layered van der waals semiconductor {Nb$_3$X$_8$ (X = Cl, I)}},
  author={Kim, Jangwon and Lee, Youjin and Choi, Young Woo and Jung, Taek Sun and Son, Suhan and Kim, Jonghyeon and Choi, Hyoung Joon and Park, Je-Geun and Kim, Jae Hoon},
  journal={ACS omega},
  volume={8},
  number={15},
  pages={14190--14196},
  year={2023},
  publisher={ACS Publications}
}

@article{huang2026suppression,
  title={Suppression of structural and magnetic phase transitions in layered exfoliated kagome semiconductor {Nb$_3$Cl$_8$}},
  author={Huang, Xinyue and Zhang, Ye and Huang, Jianqi and Ma, Zhijie and Zhu, Chenjie and Gao, Yuchen and Yang, Shiqi and Feng, Baojie and Shi, Youguo and Weng, Hongming and others},
  journal={Newton},
  volume={2},
  number={1},
  year={2026},
  publisher={Elsevier}
}

@article{ross2013electrical,
  title={Electrical control of neutral and charged excitons in a monolayer semiconductor},
  author={Ross, Jason S and Wu, Sanfeng and Yu, Hongyi and Ghimire, Nirmal J and Jones, Aaron M and Aivazian, Grant and Yan, Jiaqiang and Mandrus, David G and Xiao, Di and Yao, Wang and others},
  journal={Nature Communications},
  volume={4},
  number={1},
  pages={1474},
  year={2013},
  publisher={Nature Publishing Group UK London}
}

@article{nonnenmacher1991kelvin,
  title={Kelvin probe force microscopy},
  author={Nonnenmacher, Manuel and O’Boyle, MP and Wickramasinghe, H Kumar},
  journal={Applied Physics Letters},
  volume={58},
  number={25},
  pages={2921--2923},
  year={1991},
  publisher={American Institute of Physics}
}

@article{melitz2011kelvin,
  title={Kelvin probe force microscopy and its application},
  author={Melitz, Wilhelm and Shen, Jian and Kummel, Andrew C and Lee, Sangyeob},
  journal={Surface Science Reports},
  volume={66},
  number={1},
  pages={1--27},
  year={2011},
  publisher={Elsevier}
}

@article{garcia1998phase,
  title={Phase contrast in tapping-mode scanning force microscopy},
  author={Garcia, R. and Tamayo, J. and Calleja, M. and Garcia, F.},
  journal={Applied Physics A},
  volume={66},
  pages={S309--S312},
  year={1998},
  publisher={Springer},
  doi={10.1007/s003390051151}
}

@article{melcher2009origins,
  title={Origins of phase contrast in the atomic force microscope in liquids},
  author={Melcher, John and Carrasco, Carolina and Xu, Xin and Carrascosa, Jos{\'e} L and G{\'o}mez-Herrero, Julio and de Pablo, Pedro Jos{\'e} and Raman, Arvind},
  journal={Proceedings of the National Academy of Sciences},
  volume={106},
  number={33},
  pages={13655--13660},
  year={2009},
  publisher={National Academy of Sciences},
  doi={10.1073/pnas.0902240106}
}

@article{sun2022observation,
  title={Observation of Topological Flat Bands in the Kagome Semiconductor $\mathrm{Nb}_3\mathrm{Cl}_8$},
  author={Sun, Zhenyu and Zhou, Hui-Bo and Wang, Jinlan and Liu, Zheng and Liu, Run-Wu and Shen, Bing and Li, Ya-Jun and Jin, Xuan and Zhang, Shuang and Liu, Chuang-Han and Hitosugi, Taro and Meng, Sheng and Feng, Baojie},
  journal={Nano Letters},
  volume={22},
  number={11},
  pages={4596--4602},
  year={2022},
  publisher={American Chemical Society},
  doi={10.1021/acs.nanolett.2c00778},
  url={https://doi.org/10.1021/acs.nanolett.2c00778}
}

@article{kresse1996efficient,
  title={Efficient iterative schemes for ab initio total-energy calculations using a plane-wave basis set},
  author={Kresse, Georg and Furthm{\"u}ller, J{\"u}rgen},
  journal={Physical review B},
  volume={54},
  number={16},
  pages={11169},
  year={1996},
  publisher={APS}
}

@article{kresse1996efficiency,
  title={Efficiency of ab-initio total energy calculations for metals and semiconductors using a plane-wave basis set},
  author={Kresse, Georg and Furthm{\"u}ller, J{\"u}rgen},
  journal={Computational materials science},
  volume={6},
  number={1},
  pages={15--50},
  year={1996},
  publisher={Elsevier}
}

\end{document}